Lateral Size and Thickness Dependence in Ferroelectric Nanostructures Formed by Localized Domain Switching

Nathaniel Ng[1,2], Rajeev Ahluwalia[1], H. B. Su[2], F. Boey[2]

[1]Materials Theory and Simulation Laboratory, Institute of High Performance Computing, 1 Science Park Road, #01-01 The Capricorn, Singapore Science Park II, Singapore 117528

[2]School of Materials Science & Engineering, Nanyang Technological University, 50 Nanyang Avenue, Singapore 639798

Abstract

Ferroelectric nanostructures can be formed by local switching of domains using techniques such as piezo-force microscopy (PFM).  Understanding lateral size effects is important to determine the minimum feature size for writing ferroelectric nanostructures.  To understand these lateral size effects, we use the time-dependent-Ginzburg-Landau equations to simulate localized switching of domains for a PFM type and parallel-plate capacitor configurations.  Our investigations indicate that fringing electric fields lead to switching via 90° domain wedge nucleation for thicker films while at smaller thicknesses, the polarization switches directly by 180º rotations.  The voltage required to switch the domain increases by decreasing the lateral size and at very small lateral sizes the coercive voltage becomes so large that it becomes virtually impossible to switch the domain.  In all cases, the width of the switched region extends beyond the electrodes, due to fringing.

*Submitted to Acta Materialia*

1. Introduction

With the current advancement in nanotechnology and device miniaturization, there is much interest in studying size effects in ferroelectrics. The properties of nano-scale ferroelectric thin films substantially differ from the bulk [1], and it has been shown that strong depolarization effects can completely suppress ferroelectricity below a critical thickness. In fact, thickness dependence of ferroelectric behavior has been well studied both theoretically and experimentally [2, 3], but not much attention has been devoted to the role of lateral size. Scanning probe techniques such as PFM allow the polarization to be switched locally and arrays of these locally switched domains are the building blocks of ferroelectric memory devices (e.g. ferroelectric capacitors and dynamic random access memory (DRAM) applications). In such devices, the main limitation on lateral width is due to fringing electric fields [4], and reduction of the lateral size is key to allowing very high-density memories to be achieved [5].

Apart from the technological interest, it is also important to understand the role of the complex long-range elastic as well as electrostatic interaction on selectively switched domains. Fringing of the electric field [4] due to edge effects as well as inhomogeneous elastic deformations will crucially influence the behavior of the locally switched domains. These long-range interactions will influence the local polarizations in the neighborhood of the domains, thereby influencing the effective properties such as the polarization vs. electric field response (P-E). For example, in a recent work, Chen et. al. [6] studied the local 180° degree switching in epitaxial PZT by PFM. The found that the 90° degree domain formation relaxes the internal stresses associated with inhomogeneous electric fields. This underscores the importance of studying the effect of finite lateral width from a theoretical point of view to understand the underlying mechanisms that play a role in local switching of the domains.

Such a study may provide a guideline to choose the optimal thickness and widths of the selectively switched domains.

The effect of lateral size in ferroelectric films was studied experimentally by Lee et al [7] and Chen et al [6], as well as analytically by Wang et al [8], and Roytburd et al [9]. The analysis by Wang and Roytburd, however, did not incorporate domains and domain wall motion. In fact, both models neglected electrostatic effects, and Wang's model also neglected elastic effects. A frame work that takes into account elastic as well electrostatic long-range interacts is a useful technique to study the domain switching under inhomogeneous electric fields. In this paper, we investigate the 2D domain evolution using an LGD (Landau-Ginzburg-Devonshire) approach with the TDGL (Time-Dependent Ginzburg-Landau) relaxational equations with both electrostatic and elastic effects, focusing on local switching mechanisms for a PFM type configuration as well as for parallel plate capacitor geometry. In fact, a TDGL model has been used to investigate PFM measurements in $BiFeO_3$ [10]. However, a detailed study on size effects and the related nucleation mechanisms was not carried out. While previous TDGL models incorporated Fourier transform techniques for their numerical solvers, we solve these equations in real space for the displacement fields as well as for the electrostatic potentials for ease of introduction of the boundary conditions. We should point out that this real space method is especially convenient to simulate the physics of localized switching in a PFM setup and the parallel plate capacitor geometry. Since our investigations focus on the role of the fringing electric fields, which arise due to the finite lateral width of the switched region, we do not introduce the conventional sources of domain nucleation such as thermal noise (except in the initial conditions), and static defects due to dipoles or free charge.

The paper is organized as follows. In section 2, we describe the TDGL model which is used in section 3 to understand localized domain switching in a configuration similar to a PFM setup along with a parallel-plate capacitor geometry. Section 4 ends the paper with a summary and discussion of the results.

2. Model

Landau theory is a mean-field framework to study phase transitions, based solely on symmetry considerations [11]. In 1954, Devonshire adapted Landau's model to ferroelectrics, whereby he described thermodynamic behavior in the $BaTiO_3$ system, which is now called the LGD approach to ferroelectrics. This was later used by Haun et al [12] to develop a thermodynamic theory of the PZT system. This model has been applied to various materials and shown to be able to explain properties such as the spontaneous polarization, the intrinsic coercive field and hysteresis loops [13], and temperature dependence.

In ferroelectrics, the total free energy of the system, $F_T$, is given by:

$$F_T = F_{Landau} + F_{Ginzburg} + F_{electrostatic} + F_{elastic}, \qquad (1)$$

where $F_{Landau}$, $F_{Ginzburg}$, $F_{electrostatic}$, and $F_{elastic}$ represent the Landau, Ginzburg, electrostatic and elastic energies respectively and may be obtained as a function of the polarizations, $P_i$, which is described as follows:

The 2D Landau-type free energy expansion of the order parameter, $P_i$, is given by:

$$\begin{aligned}F_{Landau}(P_x, P_y) &= \frac{1}{2}\alpha_0(T - T_c)(P_x^2 + P_y^2) + \frac{1}{4}\alpha_{11}(P_x^4 + P_y^4) + \frac{1}{2}\alpha_{12}P_x^2 P_y^2 \\ &+ \frac{1}{6}\alpha_{111}(P_x^6 + P_y^6) + \frac{1}{2}\alpha_{112}(P_x^4 P_y^2 + P_y^4 P_x^2)\end{aligned} \qquad (2)$$

Equation (2) is a symmetry allowed expansion in the polarizations and the coefficients $\alpha_0$, $\alpha_{11}$, $\alpha_{12}$, $\alpha_{112}$, and $\alpha_{111}$ are phenomenological coefficients that can in principle be obtained from experimental data or first principles calculations [11]. Equation (2) applies to homogeneous systems, and is used to describe the phase transitions as a function of temperature. Inhomogeneous systems may be investigated by introducing an energy cost associated with creating domains in the system, i.e. the domain wall energy:

$$F_{\text{Ginzburg}} = \frac{1}{2} K \left( \left| \vec{\nabla} P_x \right|^2 + \left| \vec{\nabla} P_y \right|^2 \right), \tag{3}$$

where $K$ is a constant related to the domain wall energy.

Electrostatic effects, due to dipoles, free charge, and the electrostatic potential, are considered by the introduction of the term:

$$F_{\text{electrostatic}} = -(\mathbf{E} \cdot \mathbf{P} + \tfrac{1}{2} \varepsilon_0 \mathbf{E} \cdot \mathbf{E}), \tag{4}$$

where the electric field, $\mathbf{E}$, is obtained from the constraint given by Gauss's Law:

$$\nabla \cdot \mathbf{D} = \nabla \cdot (-\varepsilon_0 \nabla \phi + \mathbf{P}) = \rho \tag{5}$$

and $\rho = 0$ in the absence of free charge.

Similarly, elastic effects are considered by adding the elastic energy given by:

$$F_{\text{elastic}} = \tfrac{1}{2} C_{11} \left( \tilde{\varepsilon}_{xx}^2 + \tilde{\varepsilon}_{yy}^2 \right) + C_{12} \tilde{\varepsilon}_{xx} \tilde{\varepsilon}_{yy} + \tfrac{1}{2} C_{44} \tilde{\varepsilon}_{xy}^2, \text{ where} \tag{6}$$

$$\tilde{\varepsilon}_{ij} = \varepsilon_{ij} - Q_{ij} P_i P_j, \tag{7}$$

and $\varepsilon_{ij} = \tfrac{1}{2} (\partial u_i / \partial x_j)$. $\tag{8}$

Here, $C_{11}$, $C_{12}$, and $C_{44}$ are the elastic constants, $Q_{11}$, $Q_{12}$, and $Q_{44}$ are the electrostrictive coefficients, $\varepsilon_{ij}$ is the strain, $u_i$ is the displacement subject to the constraint of mechanical equilibrium given by

$$\nabla \cdot \boldsymbol{\sigma} = \sigma_{ij,j} = 0, \tag{9}$$

where the stress, $\sigma_{ij}$, is given by $\sigma_{ij} = \delta F_{elastic}/\delta \varepsilon_{ij}$.

The kinetics of the LGD approach are investigated with the time-dependent Ginzburg-Landau (TDGL) relaxational equation (also called the Allen-Cahn relaxational equation) [14] which is,

$$\frac{\partial P_i}{\partial t} = -\Gamma \frac{\delta F_T}{\delta P_i}, \tag{10}$$

where $t$ is time, and $\Gamma$ is a constant related to the domain wall mobility.

These equations have been successfully used to simulate domain formation and switching in ferroelectrics [14]

3. Simulation

These TDGL equations are rescaled for numerical convenience. In rescaled units, we put $P'_x = P_x / P_0$, $P'_y = P_y / P_0$, for the polarizations, where $P_0$ is set to the spontaneous polarization of the material, i.e. 0.647 C/m$^2$ for Pb(Zr$_{0.3}$Ti$_{0.7}$)O$_3$. The length scale is rescaled as $x' = x / \delta$, $y' = y / \delta$, where $\delta$ is chosen such that the simulation widths agree with the experimentally observed domain wall widths. We set $\phi' = \phi / \phi_0$, where $\phi_0 = \theta[(P_0\delta)/\varepsilon_0]$, for the electrostatic potential, where $\theta$ is a dimensionless parameter, arbitrarily chosen as 1.0. The elastic constants are rescaled as $C'_{ij} = C_{ij} / C_{11}$, while the electrostrictive constants are rescaled as $Q'_{ij} = Q_{ij} / Q_{11}$. For the strains, we use the rescaling, $\varepsilon'_{ij} = \varepsilon_{ij} / \varepsilon^0$, and $\tilde{\varepsilon}'_{ij} = \tilde{\varepsilon}_{ij} / \varepsilon^0$, where $\varepsilon^0 = Q_{11}P_0^2$, and for the stresses, the rescaling is $\sigma'_{ij} = \sigma_{ij} / \sigma_0$, where $\sigma_0 = C_{11}\varepsilon^0 = C_{11}Q_{11}P_0^2$.

Hence the TDGL equations may be written in rescaled form as follows:

$$\frac{\partial P'_x}{\partial t} = \Gamma|\alpha_1| \begin{bmatrix} -\frac{\alpha_1}{|\alpha_1|}P'_x - \frac{\alpha_{11}P_0^2}{|\alpha_1|}P'^3_x - \frac{\alpha_{12}P_0^2}{|\alpha_1|}P'_x P'^2_y - \frac{\alpha_{111}P_0^4}{|\alpha_1|}P'^5_x \\ -\frac{\alpha_{112}P_0^4}{|\alpha_1|}\left(P'_x P'^4_y + 2P'^3_x P'^2_y\right) + \frac{K}{|\alpha_1|\delta^2}\nabla'^2 P'_x \\ -\frac{\theta}{\varepsilon_0|\alpha_1|}\phi'_{,x'} - \frac{C_{11}Q_{11}^2 P_0^2}{|\alpha_1|}\left[2(\sigma'_{xx} + Q'_{12}\sigma'_{yy})P'_x + Q'_{44}\sigma'_{xy}P'_y\right] \end{bmatrix} \quad \text{...(11)}$$

$$\frac{\partial P'_y}{\partial t} = \Gamma|\alpha_1| \begin{bmatrix} -\frac{\alpha_1}{|\alpha_1|}P'_y - \frac{\alpha_{11}P_0^2}{|\alpha_1|}P'^3_y - \frac{\alpha_{12}P_0^2}{|\alpha_1|}P'^2_x P'_y - \frac{\alpha_{111}P_0^4}{|\alpha_1|}P'^5_y \\ -\frac{\alpha_{112}P_0^4}{|\alpha_1|}\left(P'_y P'^4_x + 2P'^3_y P'^2_x\right) + \frac{K}{|\alpha_1|\delta^2}\nabla'^2 P'_y \\ -\frac{\theta}{\varepsilon_0|\alpha_1|}\phi'_{,y'} - \frac{C_{11}Q_{11}^2 P_0^2}{|\alpha_1|}\left[2(\sigma'_{yy} + Q'_{12}\sigma'_{xx})P'_y + Q'_{44}\sigma'_{xy}P'_x\right] \end{bmatrix} \quad \text{...(12)}$$

where the constant $\Gamma|\alpha_1|$ may be absorbed into the time scale [15].

In a similar manner, Gauss's Law is rescaled as:

$$(P'_{x,x'} + P'_{y,y'}) - \theta(\phi'_{,x'x'} + \phi'_{,y'y'}) - \rho' = 0, \quad \text{...(13)}$$

and the equilibrium equation is rescaled as:

$$\begin{cases} \sigma'_{xx,x'} + \sigma'_{xy,y'} = 0 \\ \sigma'_{xy,x'} + \sigma'_{yy,y'} = 0 \end{cases} \quad \text{...(14)}$$

The rescaled TDGL equations (11)-(12) and the constraints in equations (13)-(14) are discretized onto uniform spatial grids (400 × 64, 256 × 64 and 256 × 16) using second-order symmetric finite differences and solved in real space using an iterative method with explicit time stepping. The boundaries are assumed to be traction-free, with zero gradients in the polarization ($P_{i,j} = 0$). The vanishing polarization gradients also imply that the bound charge is assumed to be perfectly compensated and hence there are no internal depolarization fields. For the electrostatic potential, we assume that $\vec{\nabla}\phi \cdot \vec{n} = 0$, where $\vec{n}$ is the unit normal to the surface, except at the electrodes where the potential is specified. We use the material parameters by M. J. Haun [12] for tetragonal Pb(Zr$_{0.3}$Ti$_{0.7}$)O$_3$ to study thin film in both a parallel-plate capacitor geometry (electrode-ferroelectric-electrode, Fig 1a) which is the building block of many devices [16] as well as a geometry somewhat approximate to that in a PFM setup (Fig 1b).

Initially, the polarizations at every point inside the grid are set to the spontaneous polarization, $P_0$, with a small initial random thermal noise (Gaussian distribution with mean $P_y/P_0 = 1.0$, standard deviation = 0.001) to ensure that any form of nucleation in the microstructure does not occur due to numerical artifacts. The system is allowed to equilibrate for 20,000 time steps to ensure a stable single domain state with $P_y > 0$.

Thereafter, the film is switched by changing the boundary condition for the electrostatic potential at the electrodes as shown in Fig 1, with $V_0 = \theta[(P_0\delta)/\varepsilon_0]\phi'_0$, where $\phi'_0 = 12.5$ for the case of film thickness, $t_{film} = 16$nm and $\phi'_0 = 200$ for $t_{film} = 64$nm, and $\omega$ is chosen such that a full hysteresis loop occurs over 800,000 time steps. The average electric field in the region below the top electrode is calculated as $E_y = -\Delta V/t_{film}$, where $\Delta V$ is the potential difference between the two electrodes.

To study lateral size effects, the width of the top electrode (Fig 1), $w$, is set as $w = 10, 32, 48, 64$ and 128 grid steps, which correspond to approximately 10, 32, 48, 64, and 128 nm, based on a grid size of $\delta = 1$nm. We set $K' = K/[|\alpha_1|\delta^2] = 1$, and $\delta = 1$nm, which sets the gradient coefficient, $K$, as $|\alpha_1|\delta^2 = 1.247 \times 10^8$ Jm$^3$C$^{-2}$. From our simulations, we have confirmed that with these choices, our domain wall thicknesses are of the same order as the experimental domain wall thicknesses of 1.3nm [17] in PZT.

We first discuss the case of $t_{film}$ = 16nm. To understand the switching mechanism, it is important to analyze the electric field distributions obtained due to the applied potentials for both the configurations. The electric field distributions in Fig 2(a) and (b) correspond to $t$ = 80,000 (and $\phi' = \phi'_0\sin(\omega t)$ = 5.67). This corresponds to the electric field, $E_y$ = -1.18$E_c$ for the PFM-type geometry and -1.14$E_c$ for the capacitor, where $E_c$ = 84.4 MV/m is the intrinsic coercive field (i.e. the electric field required to homogeneously reverse all the dipoles which may be obtained from the condition, $\partial F_T/\partial P_x = 0$ for homogeneous stress free systems). Note that at this electric field the domain is still not switched although fringing of the electric field is evident for both the cases. The fringing electric fields form due to the discontinuity in the boundary conditions, where $\phi_{,y} = 0$ at the top and bottom boundaries except at the electrodes where the boundary condition is of the form $\phi' = \phi_0'\sin(\omega t)$.

The domain structure of the film for both geometries before and after switching is as shown in Fig 3, with upward polarization vectors plotted as green, and downward polarization vectors plotted as blue: For both these cases, we find that the polarization vector changes via a 180° reversal (with the absence of 90° domain walls) regardless of geometry, and regardless of the value of $w$. Thus at this thickness, the fringing observed in Fig. 2 does not lead to formation of intermediate 90º domains for both the cases.

Figure 4 shows the fringing electric fields for both the capacitor and PFM-like geometry for the thicker 64nm film. As in figure 2, this corresponds to an electric field just before switching starts. It is clear that the fringing effects observed here are more prominent and extend to a region far beyond the region over which the potential is applied. Figures 5 and 6 show that this fringing strongly influences the switching process and leads to switching via two 90° rotations of the polarization vector, for both geometries. We should point out that since we did not artificially introduce defects into the simulation, this intermediate domain formation is purely due to the fringing electric fields.

For the PFM geometry, the domain evolution for the case $t_{film}$ = 64nm, $w$ = 32nm is as shown in Fig 5. The corresponding polarization-electric field (P-E) hysteresis loops, as well as the strain-electric field butterfly loops are also shown. Here, $<P_y/P_0>$ is calculated by averaging the polarization in the region below the top electrode. The domain structures in Fig 5 are plotted using a color scheme in which the polarizations point downwards in the blue regions, upwards in the green regions, left in the white regions, and right in the red regions, while regions of zero polarization appear as black. Hence, interfaces between a red/white region and a blue/green region indicate the presence a 90° domain wall (polarization vectors oriented at 90° to each other), while interfaces between a green and blue region, or red and white region indicates a 180° domain wall (polarization vectors oriented at 180° to each other). The red and black electrodes in Fig 1 are plotted as horizontal lines at the top & bottom of the domain structures (A) to (P) in the leftmost and rightmost columns of Fig 5. The simulation starts with a single domain state with polarization vector pointing upwards. This corresponds to point (A) of Fig 5. As the switching progresses, fringing of the electric fields leads to 90° domain wedge nucleation at point (B). These domain wedges grow in length and width at (C), and allow a second 90° rotation of the polarization vector, resulting in the appearance of blue regions (with polarization vector pointing downwards) in (D). These regions grow at the expense of other regions until the domain is completely switched (H). In the reverse switching process, nucleation of inclined domain wedges occurs (J), and these wedges grow until they merge with the 180° domain walls (K), leading to the collapse of these walls (L), (M), and further increase of the electric field leads to disappearance of the remaining domain structures (N) & (O), until the reverse switching is complete (P). The inclined 90° elastic domains at point (G) and (O), and later in point (P) of Fig 6, are somewhat similar to those observed by Chen et al [6] in their PFM setup, except that the inclined wedges do not extend completely through the film thickness.

In the case of the ferroelectric capacitor geometry, the domain evolution is shown in Fig 6. The polarization-electric field curve as well as the strain-electric field curve are obtained in the same manner as before in Fig 5. As in the case of the PFM-like geometry, nucleation of inclined 90° domain wedges occurs at point (B). However, unlike the PFM geometry case, wedges appear from the top as well as bottom electrodes. As the switching proceeds, these wedges merge (C) and grow (D). The blue reverse-switched regions start to appear at point (D) and grow at the expense of the other regions until the domain is completely reversed (I). Reversal of the electric field again leads to nucleation of inclined 90° domain wedges at the point of discontinuity between the electrode edges. The growth of these wedges (K) leads to the collapse of 180° domain walls (L), (M), (N). Green regions appear at (M) and grow at the expense of other regions until point (P).

We performed simulations for several values of the width $w$, whereby we observed results very similar to the results for $w = 32$nm with the capacitor (Fig 5) as well as the PFM-like geometry (Fig 6). However, it is instructive to compare the switching loops as a function of $w$ as well as the film thickness $t_{film}$ for both the PFM case as well as the parallel plate capacitor. The hysteresis loops are computed using the same method as in Fig 5. The results for the two different thicknesses, $t_{film} = 16$nm and $t_{film} = 64$nm are shown in Figs 7 & 8 for the PFM as well as the parallel plate capacitor respectively. It is clear from these figures that the coercive field increases for decreasing values of the lateral size, $w$. This can also be seen by replotting the coercive field data from Figs 7 & 8 as shown in Fig 9.

In all cases, at very small lateral sizes (i.e. $w = 10$nm) where the electric field becomes highly inhomogeneous, the coercive field becomes much larger than the intrinsic coercive field, $E_c$, implying there is a minimum lateral size whereby the film becomes very difficult or virtually impossible to switch. This is due to increased fringing effects as well as the additional energy costs incurred from switching regions beyond the width of the electrode. Our lateral sizes of about 10nm are close to a similar result by Wang et al [8] who proposed a minimum critical width of 7.4nm for $PbTiO_3$ and 9.1nm for $BaTiO_3$. At the other extreme, when the lateral size becomes much larger than the size of these domain wedges, the coercive field appears to approach that of the intrinsic coercive field (Fig 9).

In the previous sections, the electric field was calculated using $E_y = -\Delta V/t_{film}$. However, it is also instructive to examine the thickness and lateral-size dependence of the switching behavior in terms of the average electric field in the entire film, $E_y^{average} = \left\langle -\frac{\partial \phi}{\partial y} \right\rangle$. This is because due to fringing, the internal electric field extends beyond the electrode. Remarkably, in Fig 10, we observe an opposite trend to that of Fig 9 where the coercive field increased with decreasing lateral size. The average electric field is related to total stored electrostatic energy in the entire film, and we would expect that lower electrostatic energy is required to switch domains with smaller lateral sizes. This is the case in Fig 10 for all the configurations, although in the 64nm PFM case, we note a slight deviation, possibly due to the effects of fringing.

Due to the fringing electric fields (Figs 2 and 4), the width of the switched area becomes larger than that of the electrode (Fig 11) in all cases. In the case of the 16nm film, we observe a lack of 90° domain structures in agreement with experimental observations showing the disappearance of such structures below a certain critical size [5, 7], and that the coercive field is always larger than that of the intrinsic (Fig 9). We note that even for the 64nm thick film, the presence of 90° domain wedges does not reduce the coercive field below that of the intrinsic.

4. Conclusions

We have investigated the localized switching of nano-scale ferroelectric domains in ferroelectric thin films using real-space TDGL simulations that take into account the long-range elastic and electrostatic fields. Using the boundary conditions on the electrostatic potential we simulated two configurations, one that mimicked a PFM setup and another that is relevant to a parallel plate capacitor. We find that at sufficiently large thicknesses, fringing of the electric field causes the nucleation of 90° inclined domain wedges when a domain is switched locally. However, these wedges do not occur in films at lower thicknesses which instead undergo switching via 180° reversals of the polarization vector. An important finding, regardless of geometry and thickness of the film is that the coercive field increases with decreasing the lateral size of the switched domain and approaches the intrinsic coercive field for the large lateral sizes. In fact, for very small lateral size, it becomes very difficult or impossible to switch the domain. Another important consequence of the fringing electric fields is that the switched region in ferroelectric thin film can extend beyond the width of the electrode, which may have consequences in the packing density for nano-scale features in ferroelectric thin film.

Acknowledgements

The authors would like to acknowledge Chen Lang and David Srolovitz for useful discussions.

Figures

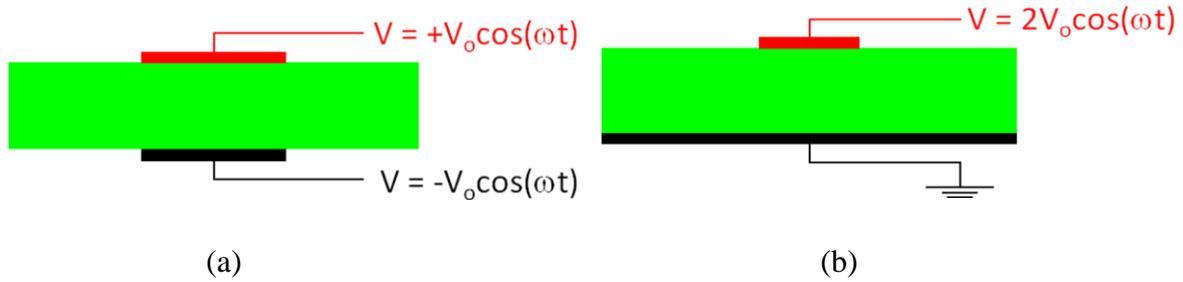

(a) (b)

Figure 1: (a) Parallel Plate Capacitor Geometry, and (b) Geometry approximate to a PFM setup.

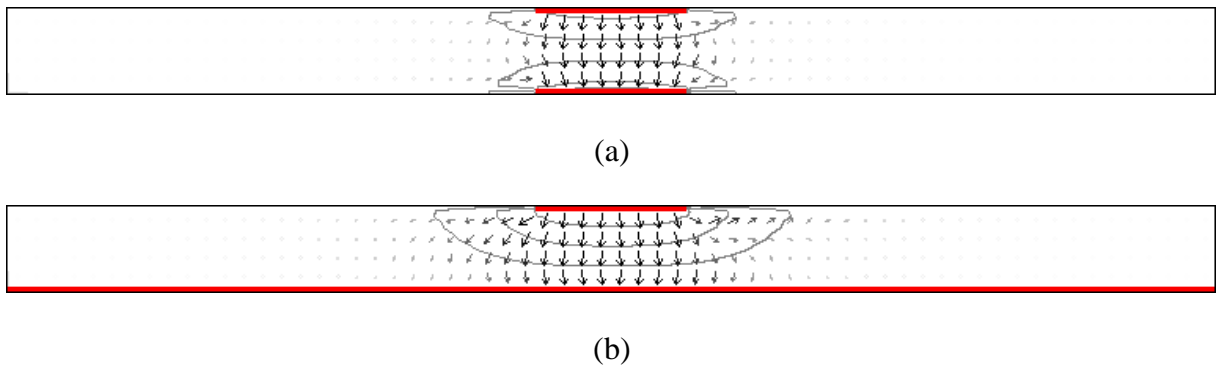

(a)

(b)

Figure 2: Electric Field for (a) the capacitor geometry and (b) the PFM-type geometry just before switching ($t = 80{,}000$).

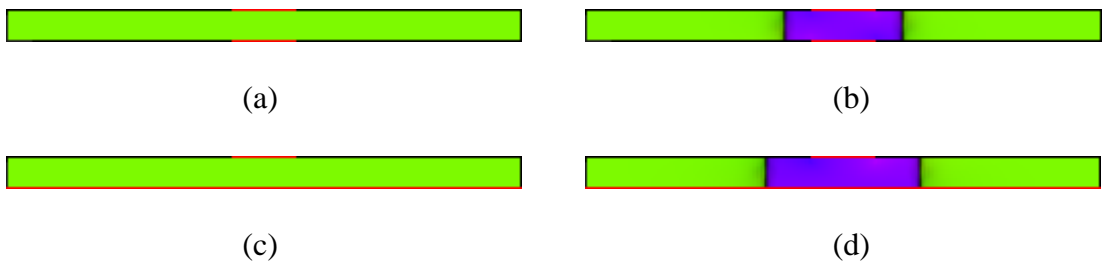

(a) (b)

(c) (d)

Figure 3: Domain structure for the geometry of Fig 1a (a) before switching ($t = 20{,}000$ time steps) and (b) after switching for $w = 32$ ($t = 300{,}000$ time steps), and for the geometry of Fig 1b, (c) before switching ($t = 20{,}000$ time steps) and (b) after switching ($t = 300{,}000$ time steps).

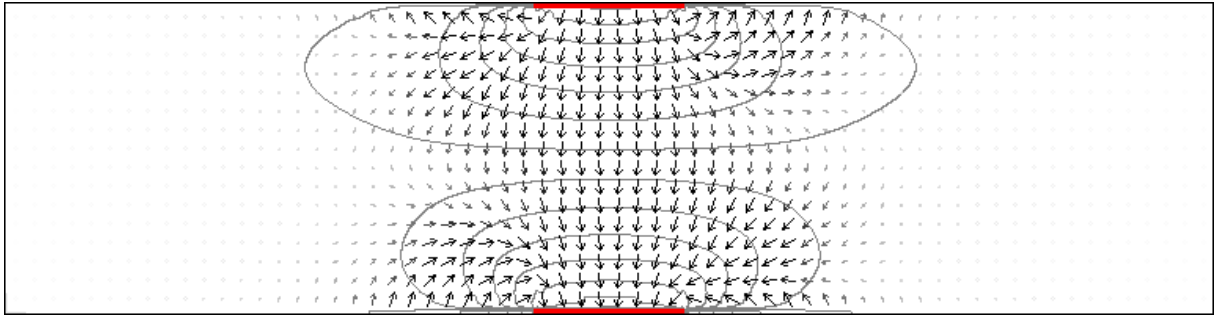

(a)

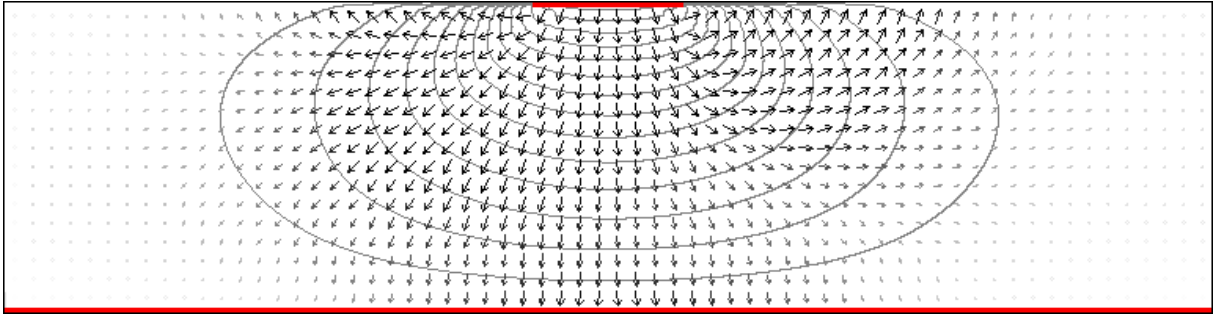

(b)

Figure 4: Electric field vectors as obtained from the solution to equation 13 for (a) the capacitor geometry (at $t = 37{,}500$, $\phi' = \phi'_0\cos(\omega t) = 27.4$) and (b) the PFM-like geometry (at $t = 30{,}000$, $\phi' = \phi'_0\cos(\omega t) = 31.4$).

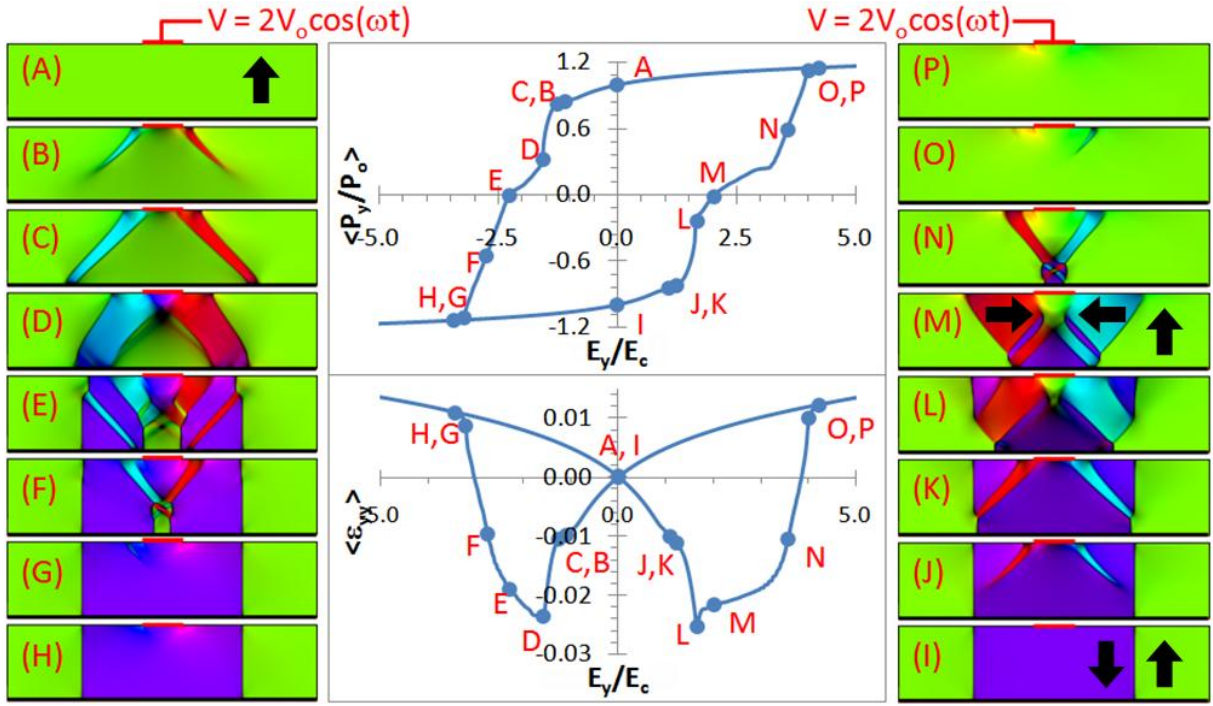

Figure 5: Domain Evolution for Film Thickness, $t_{film}$ = 64nm, and Lateral Size, $w$ = 32nm. Domain structures (A) to (P) correspond to the points (A) to (P) marked on the Hysteresis Loop ($<P_y/P_o>$ vs $E_y/E_c$) and Butterfly Loop ($<\varepsilon_{yy}>$ vs $E_y/E_c$). Black arrows refer to the polarization directions.

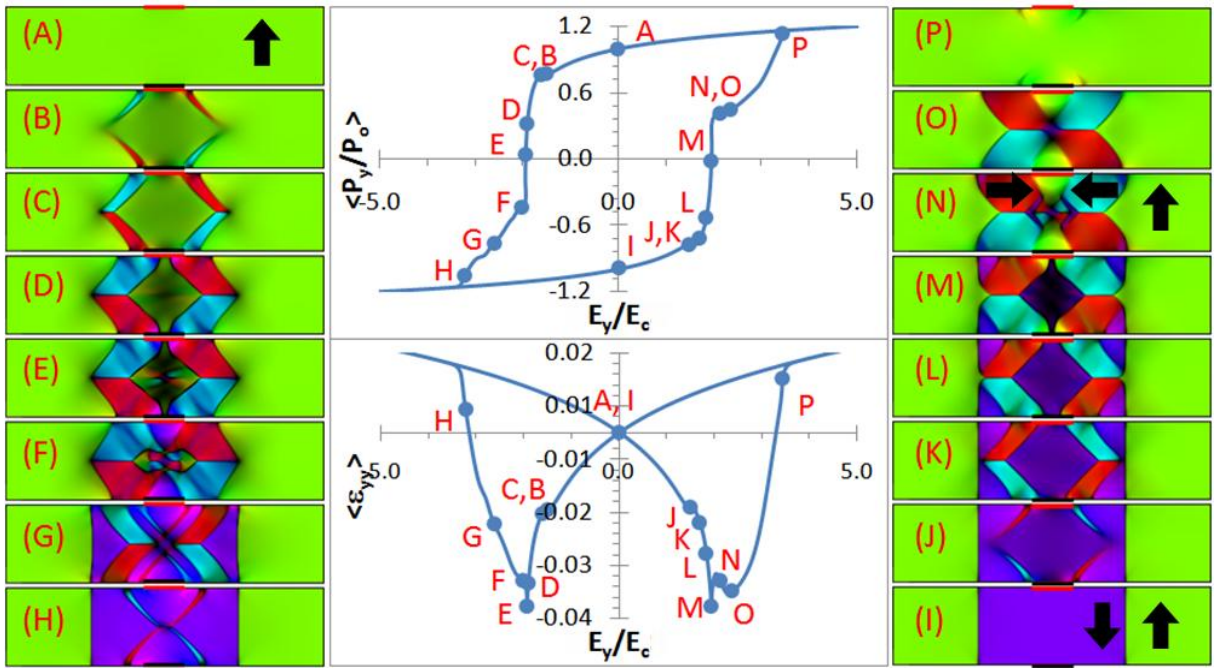

Figure 6: Domain Evolution for Capacitor Geometry with film thickness, $t_{film}$ = 64nm, and Lateral Size, $w$ = 32nm. Domain structures A to P correspond to the points A to P in the hysteresis and butterfly loops.

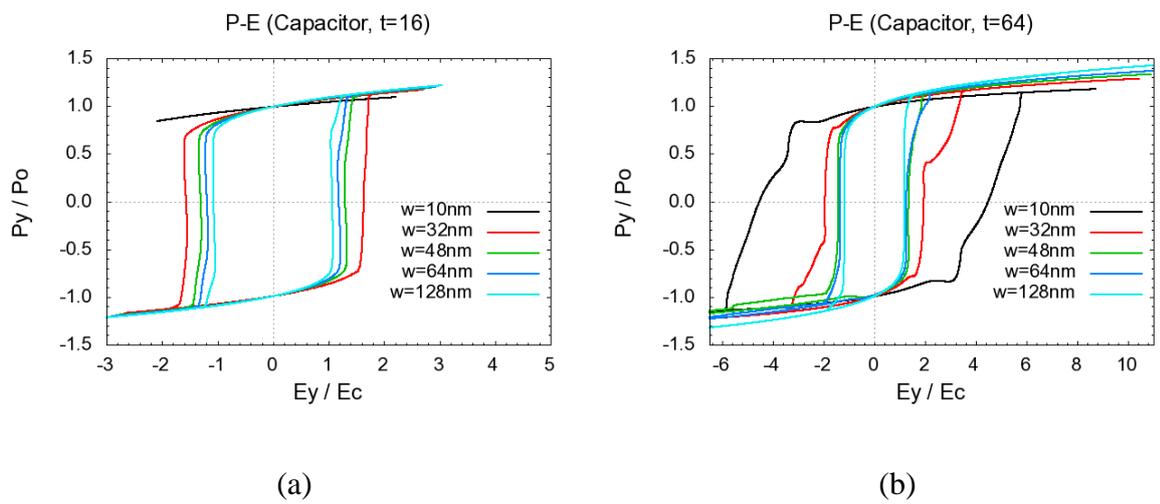

(a)          (b)

Figure 7: Average polarization versus electric field in rescaled units for film thicknesses, $t_{film}$ = 16, 64nm and lateral size, $w$ = 10, 32, 48, 64, and 128 nm in the capacitor geometry

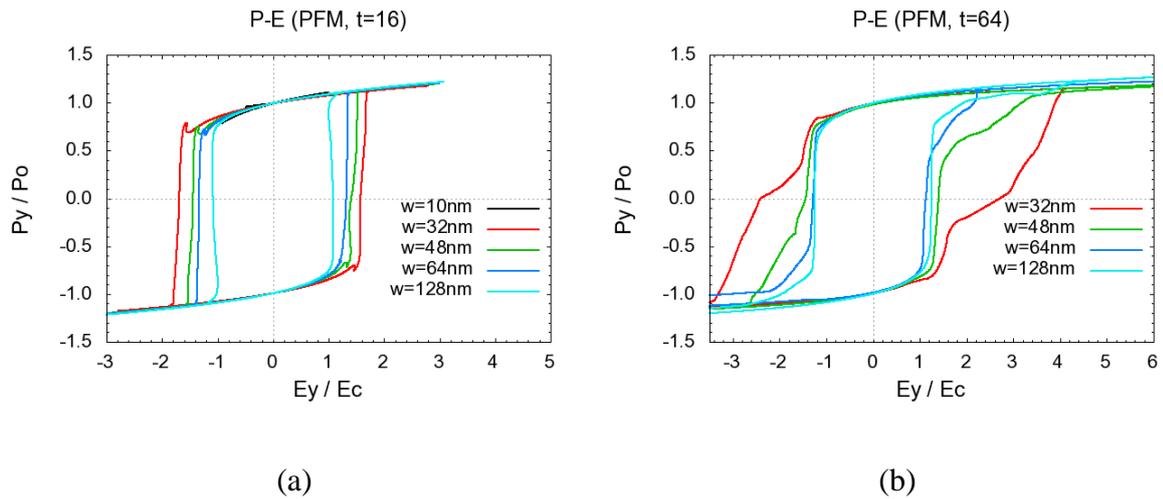

Figure 8: Average polarization versus electric field in rescaled units for film thicknesses, $t_{film}$ = 16, 64nm and $w$ = 10, 32, 48, 64, and 128 nm in the approximate PFM geometry.

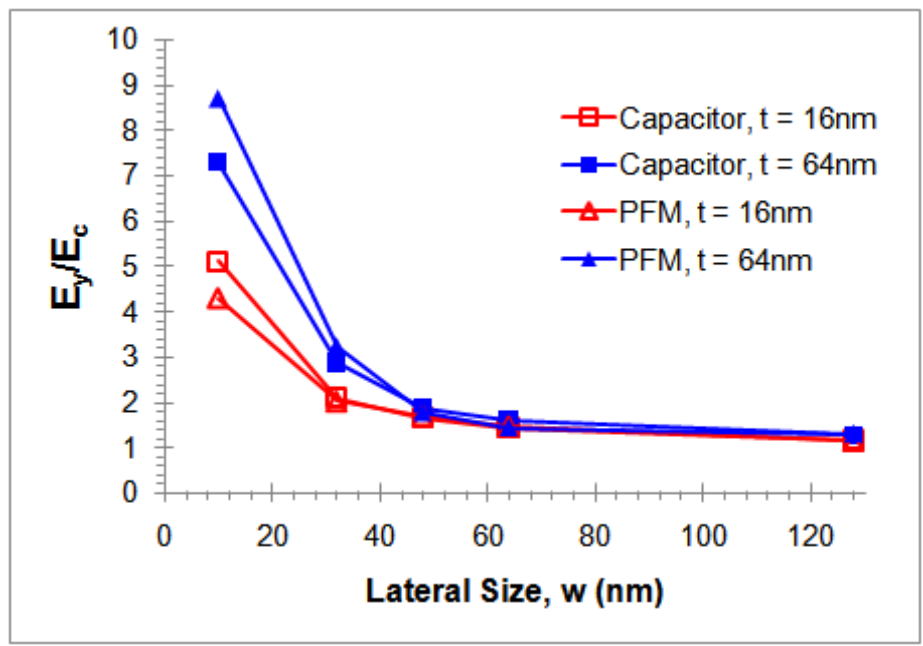

Figure 9: Rescaled electric field, $E_y / E_c$ as a function of lateral size, $w$ = 10, 32, 48, 64, and 128 nm, at different thicknesses.

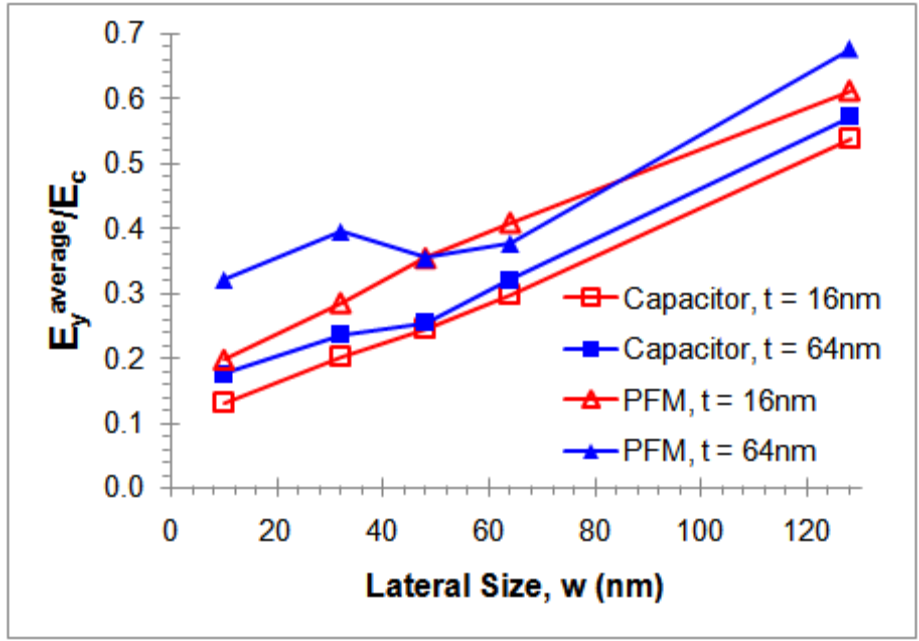

Figure 10: Rescaled electric field, $E_y^{average}/E_c$ (electric field averaged over the whole film) as a function of lateral size, $w = 10, 32, 48, 64,$ and $128$ nm, at different thicknesses.

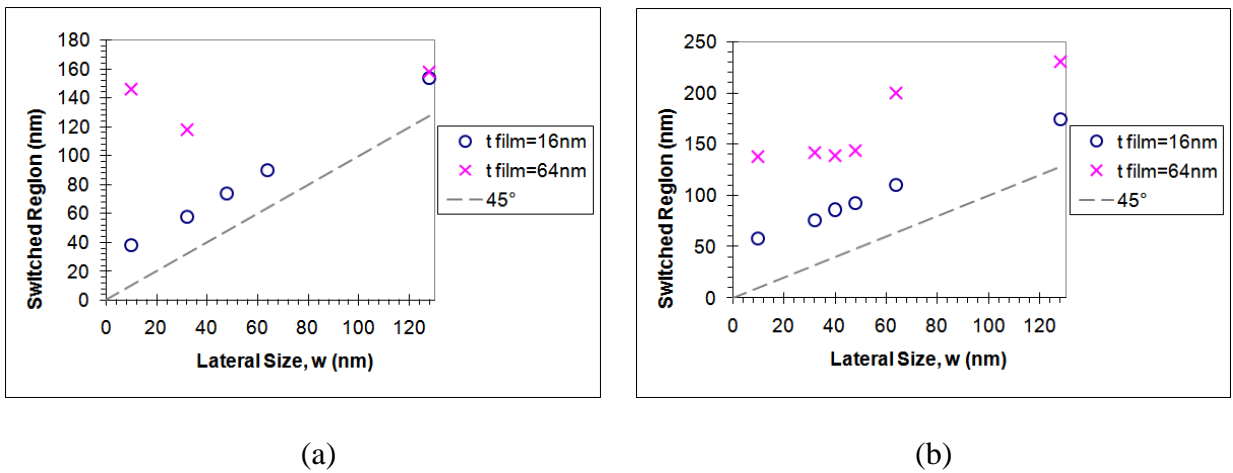

(a)                           (b)

Figure 11: Size of the region switched by the electric field for (a) Capacitor Geometry & (b) PFM Geometry, as a function of the lateral size, $w$.